# Effects of Divalent Alkaline Earth Ions on the Magnetic and Transport Features of $La_{0.65}A_{0.35}Mn_{0.95}Fe_{0.05}O_3$ Compounds

Wiqar Hussain Shah[1,2], Gul Rehman[3], S.K. Hasanain[3]

[1]Department of Physics, College of Science, King Faisal University, Hofuf, 31982, Saudi Arabia

[2]Department of Physics, Federal Urdu University, Islamabad, PAKISTAN

[3]Department of Physics, Quaid-i-Azam University, Islamabad, Pakistan

## Abstract:

We have studied the magnetic and transport behavior of doped $La_{0.65}$ $A_{0.35}$ $Mn_{0.95}Fe_{0.05}$ $O_3$ (A= Ca, Sr, Pb, Ba) compounds. All the compositions show ferromagnetic/metal to paramagnetic/insulator transition except the Pb doped sample which is insulating and ferromagnetic in the entire temperature range. The simultaneous occurrence of ferromagnetism and insulating behavior in Pb doped compound is most likely due to the presence of FM clusters separated by Fe and Mn ions that are coupled AFM and hence prevent the current from crossing the inter-domain region. The magnetization and $T_c$ are decreased by decreasing the Cation size on La site. We observed that (for a fixed Fe content of 5%) the transition temperature and magnetic moment at 77 K is a maximum for Sr doped sample and is decreasing if we increase or decrease the cation size from Sr size. The maximum value of $T_c$ and magnetic moment for Sr based sample is most likely due to the closer ionic sizes of La and Sr as compared to the other dopants (Ca, Pb, Ba). The decrease in the transition temperature and magnetic moment is due to the deviation of Mn-O-Mn angle from $180^0$ caused by the size mismatch of the A site cations. This deviation leads to weakening of the FM double exchange (DE). We observed a spin freezing type effect in the Pb doped sample below 120 K in resistivity, AC susceptibility and in magnetization. This suggests that the AFM interactions introduced by the Fe are most effective in the Pb doped composition leading to increased competition between the FM and AFM interactions. This FM and AFM interaction generates some degree of frustration leading to the appearance of spin glass like phase whose typical magnetic behavior is studied for small ion when the metallic like behavior is lost.

Corresponding Author: wiqarhussain@yahoo.com

# I. Introduction

Doped manganese oxides of the formula $R_{1-x}A_xMnO_3$ (R is rare earth ion and A is a divalant ions such as Ca, Sr, Pb or Ba) have stimulated considerable scientific and technological interest due to their diverse physical properties of X-ray [1], photon sensitivity [2], charge ordering-disordering phenomenon [3], orbital ordering [4], large magnetic entropy [5] and intrinsic colossal magnetoresistance [6]. These Physical properties of perovoskite manganites $R_{1-x}A_xMnO_3$ are known to strongly depend on doping concentration [7], oxygen stiochiometry [8], average A site ion size [9], applied external hydrostatic pressure [10] and electronegativit [11]. These properties have been examined within the frame work of double exchange model [12]. Doping the insulating $LaMnO_3$ materials, in which only $Mn^{3+}$ ion exists, with divalant ions (Ca, Sr etc) causes the conversion of a proportional number of $Mn^{3+}$ to $Mn^{4+}$. Because of Hund's coupling the electronic configuration are $Mn^{3+}$ ( $t_{2g}^3$ $e_g^1$) and $Mn^{4+}$ ($t_{2g}^3$ $e_g^0$). The presence of $Mn^{4+}$due to doping enables the $e_g$ electron of $Mn^{3+}$ ion to hop to neighboring $Mn^{4+}$ ion via DE which mediate ferromagnetism and conduction. It has recently been established that transition temperature of FM ordering of magnetic spin is also associated with a first order metal-insulator transition and extremely sensitive to lattice strain produced in the form of chemical pressure which is varied by substituting different ions for La [13]. Hwang et al. [13] have studied the doped $LaMnO_3$ with fixed carrier concentration (7:3) reveals a direct relationship between the Curie temperature $T_c$ and the average ionic radius of La site $<r_A>$. Hwang et al. [10] also studied the effects of hydrostatic pressure on the magnetoresistance in the doped $LaMnO_3$ at fixed doping level. In all cases the application of external pressure monotonically increases the Curie temperature and this was compared with the application of internal pressure which is varied by substituting different rare-earth ions for La. Thomas et al. [15] studied the transport properties of $Sm_{0.70}A_{0.30}MnO_3$ ($A=Ca^{2+}, Sr^{2+}$, $Ba^{2+}, Pb^{2+}$) and they concluded that the decrease of both $T_c$ and $T_p$ with decreasing lanthanide radius and increasing mass depends on the nature of substituent divalent cation, and they predicted that the data cannot be scaled on a universal curve of Hwang [13] as a function of average ionic radius or tolerance factor. If A sites are occupied by smaller cation (t<1) then the bending of Mn-O-Mn bonds due to the steric lattice distortion in general will result in a superexchange interaction between Mn ions and weakens the DE, which usually leads to a complex magnetic structure such as the canted spin structure or mictomagnetism structure [16]. In spite of these observation Garcia et al. [17] have investigated the magnetic frustration in mixed valence manganites. They observed the magnetotransport properties of $La_{2/3}$ $A_{1/2}$ $MnO_3$ (A=Ca, Sr) oxides reveals a gradual increase of magnetic moment with bending of Mn-O-Mn bonds. The relative strength of competing magnetic interaction is controlled not only by $R_0$ ($R_0$ is the mean radius of the ions at the La, A site) but also by electronegativity of the divalent cation. They also observed that the $T_c$ is not universal but it is sensitive to the alkaline ions.

The magnetic and transport properties of rare earth manganites also depend on the substitution on Mn site i.e. partially replacing Mn by other transition element such as Fe, Co, Cr, Ni etc [18]. Simopoulos [20] have studied the effect of Fe doping on the compound $La_{1-x}Ca_xMnO_3$ by means of Mossbauer spectroscopy and observed that Fe is couple antiferromagnetically to its neighbors. Ogal [21] studied the effect of the doping in the compound $La_{0.70}Ca_{0.30}Mn_{1-x}Fe_xO_3$ and observed the occurrence of a localization-delocalization transition in the system at critical concentration. Ahn et al. [22] have studied the effect of Fe doping on the Mn site in the FM and AFM phases of $La_{1-x}Ca_xMnO_3$ upon doping, no appreciable structure changes have been found in either series. However, conduction and ferromagnetism have been consistently suppressed by Fe doping and colossal magnetoresistance has been shifted to lower temperatures, and in some cases enhanced by Fe doping. It has been shown [23, 24] that Fe and Co doping leads to the formation of locally anti-ferromagnetically coupled spins or clusters with localized spin excitation. Leung and Morish [25] have shown that the Fe doped $La_{1-x}Pb_xMn_{1-y}Fe_yO_3$ have both Fe-O-Mn and Fe-O-Fe super exchange play a role in determining the magnetic structure. Hasanain et al. [26] studied the effect of Fe doping on the transport and magnetic behavior in $La_{0.65}Ca_{0.35}Mn_{1-y}Fe_yO_3$ and observed that the variation in the critical temperature $T_c$, confinement length, activation energy, magnetic moment and magnetoresistance show a rapid change at 4-5 % Fe. They suggested that the effect of Fe is seen to be consistent with the disruption of the Mn-Mn exchange possible due to the formation of antiferromagnetic magnetic clusters.

Here we present first time an extensive study of the doping on La site, that how the chemical pressure produced by different alkaline earth ions (Ca, Sr, Pb, and Ba) affects the magnetic moments, magnetic spin excitation, magnetic phase transformation, resistivity, localization length and other magnetic and transport properties, e.g. relationship between the Curie temperature $T_c$ and the average ionic radius $\langle r_A \rangle$.

## II. Experiments:

All samples reported in the present study were synthesized by standard solid state reaction procedure. Stoichiometric compositions of $La_{0.65}A_{0.35}Mn_{0.95}Fe_{0.05}O_3$ (A= Ca, Sr, Pb, and Ba), were prepared by mixing the equimolar amounts of $La_2O_3$, $CaCO_3$, $SrO_2$, $PbO$, $BaO_2$, $Mn_2O_3$ and $Fe_2O_3$ (having +99.99% purity). Powder of these oxides and carbonate were mixed thoroughly in acetone and were finely ground in an electric grinder for thirty minutes. After drying, the mixtures were calcined in alumina boats at 1000 °C for 16 hours, then cooled to room temperature, reground and again heated at 1100 °C for 17 hours. Following cooling to room temperature, they were reground and again heated at 1200 °C for 17 hours. After the third heat treatment, the materials were ground to fine powder and were pressed in to pellets of 13 mm diameter and 2 mm thickness under a pressure of 5 tons/inch$^2$. These pellets were heated at 1250 °C for 17 hours for the final heat treatment. X-ray diffraction (XRD) measurements were carried out to confirm that single phase materials had been prepared.

The XRD data of $La_{0.65}A_{0.35}Mn_{0.95}Fe_{0.05}O_3$ at room temperature were collected by step scanning over the angular range of $15° \le 2\theta \le 70°$ as shown in **Fig. 1**. All peaks were successfully indexed on the basis of a tetragonal unit cell. The lattice parameters, which were obtained using least square fitting procedures are, **a** = 5.460(2) Å and **c** = 7.730(9) Å. From the XRD data as shown here no structural changes were observed with the doping of different cations. No impurities or secondary phases were observed.

Four probe techniques was used for the resistivity measurements, with temperature varying from 77 to 300 K. AC susceptibility studies were conducted using a self-made ac probe with a split secondary (astatically wound) and a commercial lock-in amplifier. AC studies were conducted in the range $h_{ac}$=5 Oe, f= 131 Hz and $0 < H_{dc} < 550$ Oe. DC magnetic fields were obtained from a homemade solenoid magnet with the dc field direction parallel to that of ac field. Studies with superposed dc fields were conducted during warm up after the sample had been cooled down to 81 K.

## III. Results and Discussion:

The samples studied in these experiments belongs to a general family of Fe doped CMR compounds that have been investigated from the point of view that how different cation of different radii change the magnetic and transport properties of these compounds substituted on La site. The ionic radii for these cations are given as1.06 Å, 1.27 Å, 1.32 Å and 1.43 Å for Ca, Sr, Pb and Ba respectively [27].

**Resistivity Measurements:**

The temperature dependence of the resistivity of these samples $La_{0.65}A_{0.35} Mn_{0.95}Fe_{0.05} O_3$ (A= Ca, Sr, Pb, Ba) have been studied through standard four probe technique. From the resistivity measurements, $La_{0.65} Ca_{0.35} Mn_{0.95}Fe_{0.05} O_3$ (LCMFe) sample has metal to insulating phase transition at temperature $T_P$ of 185 K as shown in **Fig. 2.** Above $T_p$ the sample has insulating like behavior, the resistivity increases with decreasing temperature, while cooling down the sample passing through the peak resistivity at 185 K the sample behave like metal as resistivity decreasing by decreasing the temperature down to 80 K. The possible explanation of this behavior can be explained by the polaron conduction [26] above $T_p$ and below $T_p$, as being due to the double exchange mediated hopping of the hole between the $Mn^{3+}$ and $Mn^{4+}$ ions. Our analysis of the resistive behavior below $T_p$ follows the study of Viret et al. [29] where he assume that the system still has a polaronic character but the activation energy tends to decrease with decreasing temperature as the magnetic correlations develop and the spontaneous moment allows the hopping to occur over lowered energy barriers. Below $T_p$ the activation energy as a function of temperature is observed to be decreasing linearly with temperature as shown in the inset of **Fig 2**. The activation energy above $T_C$ is 145 meV that is in good agreement with the reported values [28]. If we compare our results with the Fe undoped sample ($La_{0.65}Ca_{0.35}MnO_3$) we note that the transition from insulting to metallic state occurs at $T_p$=270 K and the activation energy above $T_c$ is 129 meV [30]. It is clear that 5% Fe doping samples lowers $T_p$ and increase the activation energy and the resistivity at the peak which is very important investigation of this study. We observed that in our system Fe doping tends to result in weakening of the double exchange interaction because Fe ions being in +3 state and not participating in the double exchange mechanism. The continuous decease of the activation energy at low temperatures coincides with the increase of the spontaneous moment in this temperature range. The $La_{0.65}Sr_{0.35}Mn_{0.95}Fe_{0.05}O_3$ shows metallic as well as insulating behavior below and above the transition temperature $T_p$. The transition from insulator to metallic region is broad extending to over 120 K and the activation energy is 41 meV. This is the lowest activation energy compared to the other samples studied here, because the Sr size is very close to the La size, compare to Ca, Pb, Ba sizes. The resistivity of $La_{0.65}$ $Pb_{0.35}$ $Mn_{0.95}Fe_{0.05}$ $O_3$ shows anomalous behaviour that there is no transition from insulator to metal and is insulating in the entire temperature region as shown in **Fig. 3**. The resistivity of this sample is increasing with decreasing the temperature and there is flat region (190 -210 K) where the resistivity is smooth but large increase in the resistivity is observed for the temperature T< 120 K which is the indication of large scattering and it may be due to the disorder of spins, which leads to very large increase in the resistivity in this temperature region. This is also an indication of enhanced magnetic scattering, where the spins, as discussed above, appear to be freeze in different orientations. It thus appears that the effect of Fe ions in the La:Ca 65:35 ratio manganites is particularly strong in the compounds with Pb as the divalent element. The unique and interesting role of Pb in these compounds cannot be attributed to size effects because Ba has even larger size than Pb. One notes however that in terms of electronegativity there is a marked difference between Pb on the one hand and the other divalent elements of this study. e.g. Ca has an electronegativity of 1, Sr has 0.95, Ba has 0.89, and Pb has an electronegativity of 2.33. This very large difference of the electronegativity of Pb suggests that it may be more electrons attracting than the other divalent elements studied here. This could in turn affect the electronic state of the Fe-Mn ions, and hence disturb the exchange interaction. Thus this composition falls in the unusual category of ferromagnetic and insulating system. Typically such behavior is observed in charge ordered manganites. The occurrence of this behavior is close to optimal doping of the divalent ions is obviously due to the special effects of Fe in this particular matrix. Comparing these result with those for Fe un-doped sample, Gutierreg et al. [31] have shown that $La_{0.70}Pb_{0.30}MnO_3$ has $T_c$=345 K and $T_P$=270 K while $La_{0.70}Pb_{0.30}Mn_{0.95}Fe_{0.05}O_3$ has $T_c$= 295 K and shows metal-insulator transition at 220 K. Thomass et al. [15] found that $La_{0.70}Pb_{0.30}MnO_3$ shows magnetic transition temperature at 300 K and they also observed that $La_{0.70}Pb_{0.30}MnO_3$ has a broad peak centered around 390 K ($T_p$=390 K) which is uncharacteristically higher than the magnetic order temperature $T_c$=300 K. Gutierrez [31] have shown that only composition of 5% and 10% Fe doping in $La_{0.70}$ $Pb_{0.30}$ $Mn_{1-x}Fe_xO_3$ exhibit magnetoresistance effect. They also observed clear insulating behavior for the sample with x= 0.2 and 0.3 even measured under 6 T of applied magnetic field. It thus appears that the effect of Fe ions (decrease of $T_c$ and absence of metallic transition) is particularly strong for the La:Ca 65-35 composition of Pb. The simultaneous occurrence of ferromagnetism and insulating behavior is most likely due to the presence of FM clusters separated by Fe or Mn ions that are coupled antiferromagnetically and hence prevent the current from crossing the inter-domain region. Measuring the resistivity of $La_{0.65}$ $Ba_{0.35}$ $Mn_{0.95}Fe_{0.05}$ $O_3$ the transition temperature is observed at 207 K. Mclorry et al. [32] have found that the Fe undoped sample ($La_{0.65}Ba_{0.35}MnO_3$) shows $T_c$= 330 K and they observed that this sample does not exhibit a metal insulator transition but remains metallic on both side of FM-PM transition while Yuang et al. [33] observed that $La_{2/3}Ba_{1/3}MnO_3$ shows thermally activated insulating like behavior at high temperature with resistivity rising on cooling. After the resistivity passing a maximum at $T_P$ ~330 K further cooling brings about a sharp reduction in resistivity, which indicates metallic behavior.

The very pronounced change in the resistivity for the Pb based very important results, which suggests that the disorder exceed a critical threshold value. It is possible that the threshold is connected with the percolation paths in the doped materials. The transport results shown above clearly demonstrate that the partial replacement of La by different divalent cation or Mn by Fe favors insulating and AF behavior, opposing the effect of double exchange. Since Fe doping is the direct replacement of $Mn^{+3}$ by $Fe^{+3}$, the experimental results suggests that the sites that are now occupied by $Fe^{+3}$ can no longer effectively participate in the DE process. The mechanism that $Fe^{+3}$ terminate the DE process arises purely from the electronic structure of the materials.

**<u>Magnetization Measurements</u>**

From the magnetization (M-H) measurements we find out the saturated magnetic moment at 77 K which is shown in **Fig 4**. Comparing the magnetic behaviour from the magnetization in 15 kOe DC magnetic field for samples $La_{0.65}$ $A_{0.35}$ $Mn_{0.95}Fe_{0.05}$ $O_3$ (A= Ca, Sr, Pb, Ba) with fixed $Mn^{3+}/Mn^{4+}$ ratio, we observed that all the sample are PM at room temperature except the Sr doped sample which is ferromagnetic. If we compare the saturated magnetic moment for our Fe doped samples with the observed magnetic moment for the un-doped samples [26], we confirm that Fe doping decrease the magnetic moment of the system.
The temperautre dependence of DC magnetization was studied in both the field cooled (FC) and zero field cooled (ZFC) condition for all samples to observe any spin freezing like behaviour. When we cool the sample in the absence of magnetic field (ZFC) the magnetic moment in the FM region reaches to saturation for LCMFe but the magnetic moment for LPMFe and LBMFe is decreasing with decreasing temperature. When the sample is cooled in 50 Oe DC magnetic field, the magnetic moment is increasing in the FM region for all the samples. The result of $La_{0.65}Pb_{0.35}Mn_{0.95}Fe_{0.05}O_3$ for DC magnetization in the field of 50 Oe in both condations is shown in **Fig 5**. Both data were obtained while warming the sample in this field. The FC magnetization is continous to increase down to the lowest temperature and we observe a transition from PM to FM phase. An interesting magnetic behavior is observed that in the ZFC condition, in the FM region the magnetic moment is decreasing with decreasing temperature. The magnetic moment in the temperature range 190-210 K is increaisng slowly with decreasing temperature in both FC and ZFC which is consistant in the resistity data and below 190 K both FC and ZFC data are seperated from one another. We observe a continuous decrease in the magnetic moment below 110 K in ZFC. There is 25% difference between the FC and ZFC magnetization at the lowest temperature. This difference decreases as the temperature is increased and becomes zero at 208 K. This behavior may be explain as at low temperature the magnetic moment of LPMFe are frozen in arbitrary state with no long range order state. Because of the spin disorder there is a strong competition between FM double exchange and AFM super exchange so that a particular spin will receive conflicting information on how to order from its neighbors and it might not possible for the system to chose certain spin configuration to minimize its energy and thus result in frustration which leads to decrease the magnetic moment.

In the temperature region 190-208 K there is a slowing down of the rise in the magnetic moment in the transition region and this behavior is also observed in the resistivity data at the same temperature range. As the temperature is decreased the strong DE magnetic interaction between Mn ions is decreasing and the super exchange interaction is increasing and there is a competition between the two phases. This FM and AFM interaction generates some degree of frustration leading to the appearance of spin glass like phase whose typical magnetic behavior is studied for small ion when the metallic like behavior is lost.

**AC susceptibility Measurements:**

The AC susceptibility measurements of the samples $La_{0.65}$ $A_{0.35}$ $Mn_{0.95}Fe_{0.05}$ $O_3$ (A= Ca, Sr, Pb, Ba) can be divided into two categories. Firstly we observe the behavior of in-phase ($\chi^{/}$) and out of phase ($\chi^{//}$) part as a function of temperature. From the in-phase part of the susceptibility as shown in **Fig. 6**, we can find out the FM Curie temperature $T_c$. It is clear from the susceptibility data that the highest $T_c$ is observed for LSMFe is due to the less size mismatch between La and Sr. We have observed the most important result that $T_c$ is decreasing with decreasing the ionic radii because decreasing ionic radii lead to the decrease of Mn-O-Mn bond angle which decrease the hopping between Mn site and results in decrease of transition temperature but in case of Sr doped sample the lattice distortion is less and the hopping between Mn site is large which lead to high $T_c$. The spin freezing like behavior is also observed in these measurements below 120 K in Pb doped sample. The in-phase and out of phase part of AC susceptibility for $La_{0.65}Pb_{0.35}Mn_{0.95}Fe_{0.05}O_3$ at frequency of 131 Hz and AC amplitude of 5 Oe is shown in **Fig 7** . The in-phase part rises sharply around 220 K and reaches to maximum around $T_c$ below which there is typical decrease of FM AC susceptibility and finally there is a sharp decrease below 120 K. The out of phase part $\chi^{//}$ ( the loss component) for $La_{0.65}Pb_{0.35}Mn_{0.95}Fe_{0.05}O_3$ has a peak close to $T_c$ and there is increase in the $\chi^{//}$ as we decrease the temperature below 120 K which shows the behavior of the spin freezing where disordering of spin leads to large out of phase part. Below this temperature the AFM interaction introduced by $Fe^{3+}$ ions begins to creat significance degree of frustration and this frustration is strong as compare to our other samples. Secondly we examine the effect of DC magnetic field (H= 550 Oe) on the response of the in-phase and out of phase part of susceptibility. The spin freezing behavior of Pb doped sample is also investigated in measurement of AC susceptibility in a superimposed DC field where we see the scattering process is minimized by this magnetic field as shown in **Fig 7**. We found both the in-phase and out of phase part are very strongly suppressed in the FM region for all the samples but large suppression of $\chi^{/}$ at low temperature was observed for Pb doped sample.

## IV. Conclusion:

We arrived at the conclusions that Fe doping weakens the double exchange and favors the anti-ferromagnetism and in one case completely suppress the metallic state. The magnetization and $T_c$ are decreased either replacing Fe on Mn site or decreasing the Cation size on La site. All the samples are PM at room temperature except Sr based sample which is FM. We observed that (for a fixed Fe content of 5%) the transition temperature and magnetic moment at 77 K is maximum for Sr doped sample and is decreasing if we increase or decrease the cation size from Sr size. The maximum value of $T_c$ and magnetic moment for Sr based sample is most likely due to the closer ionic sizes of La and Sr as compared to the other dopants (Ba, Pb, Ca). The decrease in the transition temperature and magnetic moment is due to the deviation of Mn-O-Mn angle from $180^0$ caused by the size mismatch of the A site cations which is leads to weakening of the FM double exchange. We observed a spin freezing type effect in the Pb doped sample below 120 K which suggests that the AFM interactions introduced by the Fe ions are most effective in the Pb doped composition leading to increased competition between the FM and AFM interactions. All the compositions below their $T_c$'s are ferromagnetic-metallic except Pb based sample which shows FM insulator behavior. The simultaneous occurrence of ferromagnetism and insulating behavior in Pb doped compound is most likely due to the presence of FM clusters separated by Fe and Mn ions that are coupled anti-ferromagnetically and hence prevent the current from crossing the inter-domain region. Rapid increase in resistivity is an indication of enhanced magnetic scattering in a temperature region where the spins, appear to be freeze in different orientations. The unique role of Pb in these compounds cannot be attributed to size effects because Ba has even larger size than Pb, one note however that in terms of electronegativity there is a marked difference between Pb on the one hand and the other divalent elements of this study. The very large value of the electronegativity of Pb suggests that it may be more electrons attracting than the other divalent elements which could in turn affect the electronic state of the Fe-Mn ions, and hence disturb the exchange interaction thereby leading to large spin scattering.

## References:

1. V. Kiryukhin, D. Casa, J.P. Hill, B. Keimer, A. Vigliante, Y. Tomioka, Y. Tokura, Nature **386** (1997) 813.
2. K. Miyano, T. Tanaka, Y. Timioka, Y. Tokura, Phys. Rev. Lett**.** **78** (1997) 4257.
3. Y. Tomioka, A. Asamitsu, Y. Moritomo, H. Kuwahara, Y. Tokura, Phys. Rev. Lett. **74** (1995) 5108.
4. Taka-hisa Arima, Kenji Nakamura , Phys. Rev. B **60** (1996) R 15013
5. Z.B. Guo, Y. W. Du, J.S. Zhu, H. Huang, W.P. Ding, D. Feng, Phys. Rev. Lett. **78** (1997) 1142.
6. G.C. Xiong, Qi Li, H. L. Ju, S.N. Mao, L. Senapati, X. X. Xi, R. L. Greene, T. Venkatesan, Appl. Phys. Lett. **66** (1995) 1427.
7. A. Urushibara, Y. Moritomo, T. Arima, Asamitsu, G. Kido, Y. Tokura, Phys. Rev. B **51** (1995) 14103.
8. H.L. Ju, J. Gopalkrishnan, J.L. Peng, Qi Li, G.C. Xiong, T. Venkatesan, R.L. Greene, Phys. Rev. B **51** (1995) 6143.
9. H.Y. Hwang, S.W. Cheong, P.G. Redaelli, M. Marezio and B. Batlogg, Phys. Rev. Lett. **75** (1995) 914.
10. H.Y. Hwang, T.T. M.Palstra, S.W. Cheong, and B. Batlogg, Phys. Rev. B **52** (1995) 15046
11. J. Fontcuberta, J. L. Garcia-Munoz, Suaaidi, B. Martinez, S. Pinol, X. Obradors, J. Appl. Phys. **81** (1997) 5481.
12. C. Zener, Physical Rev. **82** (1951) 403.
13. H.Y. Hwang, S. W. Cheong, P. G. Redaelli, M. Marezio and B. Batlogg, Phys. Rev. Lett. **75** (1995) 914.
14. Zaibing Guo, Jianronng Zhang, Ning Zhang, Weiping Ding, He Hung, and Youwei Du, Appl. Phys. Lett. **70** (1997) 1897.
15. R.M. Thomas, L. Ranno, and J. M.D.Coey, J. Appl. Phys. **81** (1997) 5763.
16. J. R. Sun, G. H. Rao, and J. K. Liang, Appl. Phys. Lett. **70** (1997) 1900.
17. J. L. Garcia. Munoz, J. Fontcuberta, B. Martinez, A. Seffar, S. Pinol and X. Obradors,

Phys. Rev. B **55** (1997) R668.

18. K. Ghosh, S.B. Ogale. R. Ramesh, R. L. Greene, T. Venkatesan, K. M. Gapchup, Ravi Bathe and S. I. Patil Phys. Rev. B **59** (1999) 533.
19. Youngsun, Xiaojun and Yuheng Zhang Phys. Rev. B **63** 054404.
20. Simopoulos et al Phys. Rev. B **59** (1999) 1263.
21. S.B. Ogale, R. Shreekala, Ravi Bathe, S. K. Date, S.I. Patil, B. Hannoyer, F. Patil, and G. Marest. Phys. Rev. B **57,** (1998) 7841.
22. K. H. Ahn, X. W. Wu, K. Liu and C. L. Chien, Phys. Rev. B. **54** (1996) 15299.
23. Chechersky V, Nath A, Isaac I, Franck J P, Ghosh K, Ju H and Greene R L , Phys. Rev. B **59** (1999) 497.
24. De Teresa J M, Ibara M R, Algarabel P A, Ritter C, Marquina C, Blasco J, Garcia J, del Moral A and arnold Z. Nature **386** (1997) 256.
25. L.K. Leung, A. H. Morish, and B.J. Evans, Phys. Rev. B **13** (1976) 4069.
26. S. K. Hasanain, M. Nadeem, Wiqar Hussain Shah, M. J. Akhtar and M. M. Hasan, J. Phys. Condens. Matter **12** (2000) 9007.
27. S. Jin, M. McCormack, and T. H. Tiefel, J. Appl. Phys. **76** (1995) 6929.
28. L. Seetha Lakshmi, V. Sridaran, D.V. Natarajan, V. Sankara Sastry and T.S. Radhakrishnan, http//arxiv.org/cond-mat/0203425.
29. M. Viret, L. Ranno, and M.D. Coey, J. Appl. Phys. **81** (1997) 4964.
30. S.K. Hasanain, Wiqar Hussain Shah, A. Mumtaz, M. Nadeem, M.J. Akhtar, Journal of Magnetism and Magnetic Materials **271** (2004) 79–87.
31. J. Gutierrez, A. Pena, J.M. Barandiaran, J.L. Pizarro, T. Hernandez, L. Lezama, M. Insausti and T. Rojo, Phys. Rev. B **61** (2000 ) 9028.
32. D.N. Mcllroy, C. Waldfried, Jaindi Zhang, J.W. Choi, F. Foong, S.H. Liou and P.A. Dowben, Phys. Rev. B **54** (1996) 17438.
33. S.L. Yuan et al, J. Phys. Condense. Matter **14** (2002) 173.
34. H.L. Ju, Y.S. Nam, J.E. Lee and H.S. Shin J. Magn. Magn. Matter **219** (2000).

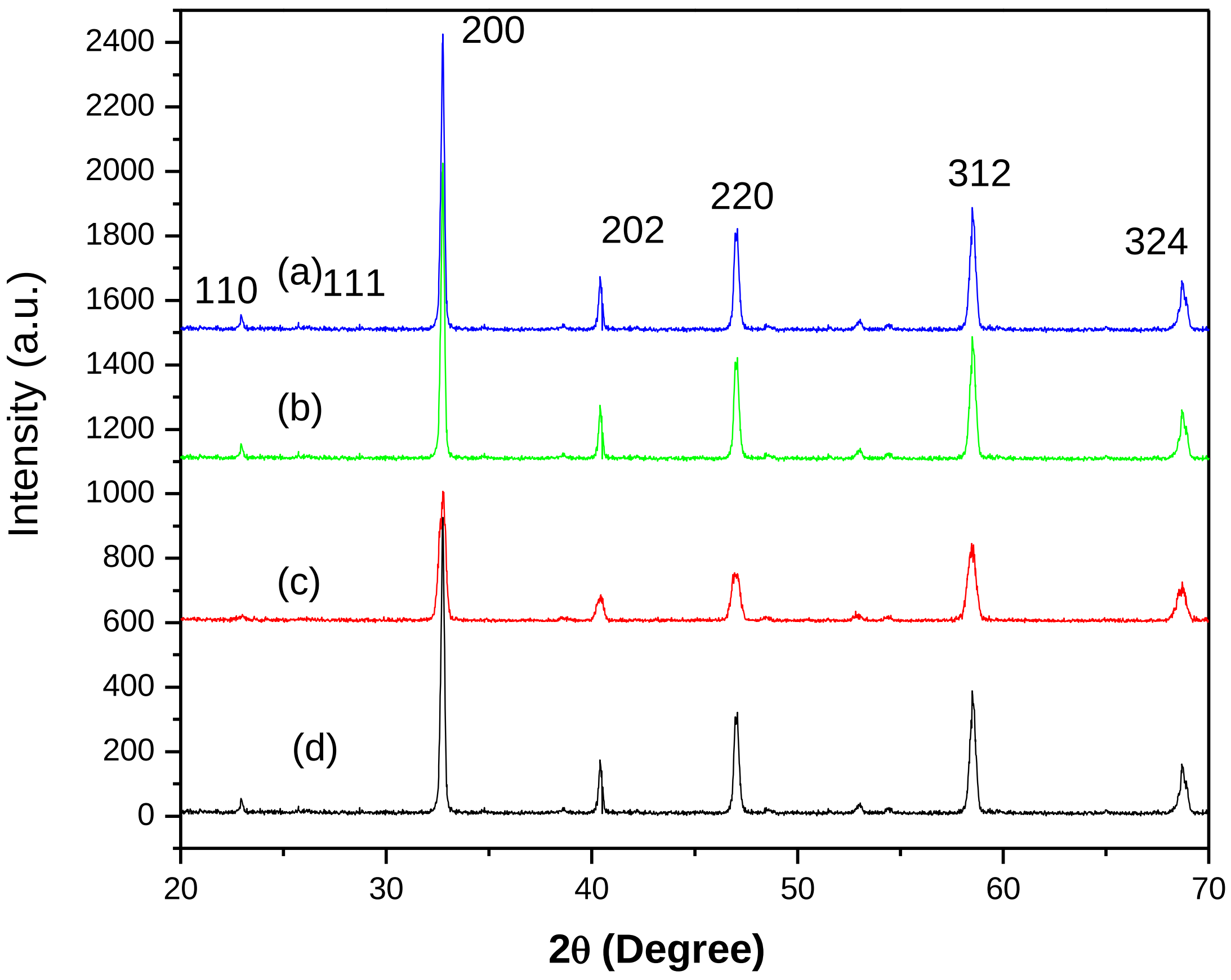

**Fig. 1**. X-ray diffraction patterns for $La_{0.65}A_{0.35}$ $Mn_{0.95}Fe_{0.05}$ $O_3$ (A= Ca, Sr, Pb, Ba), are shown, where curve (a), (b), (c) and (d) represents the Ca, Sr, Pb and Ba ((a) A=Ca; (b) A=Sr; (c) A=Pb, (d) A=Ba,) doped samples respectively.

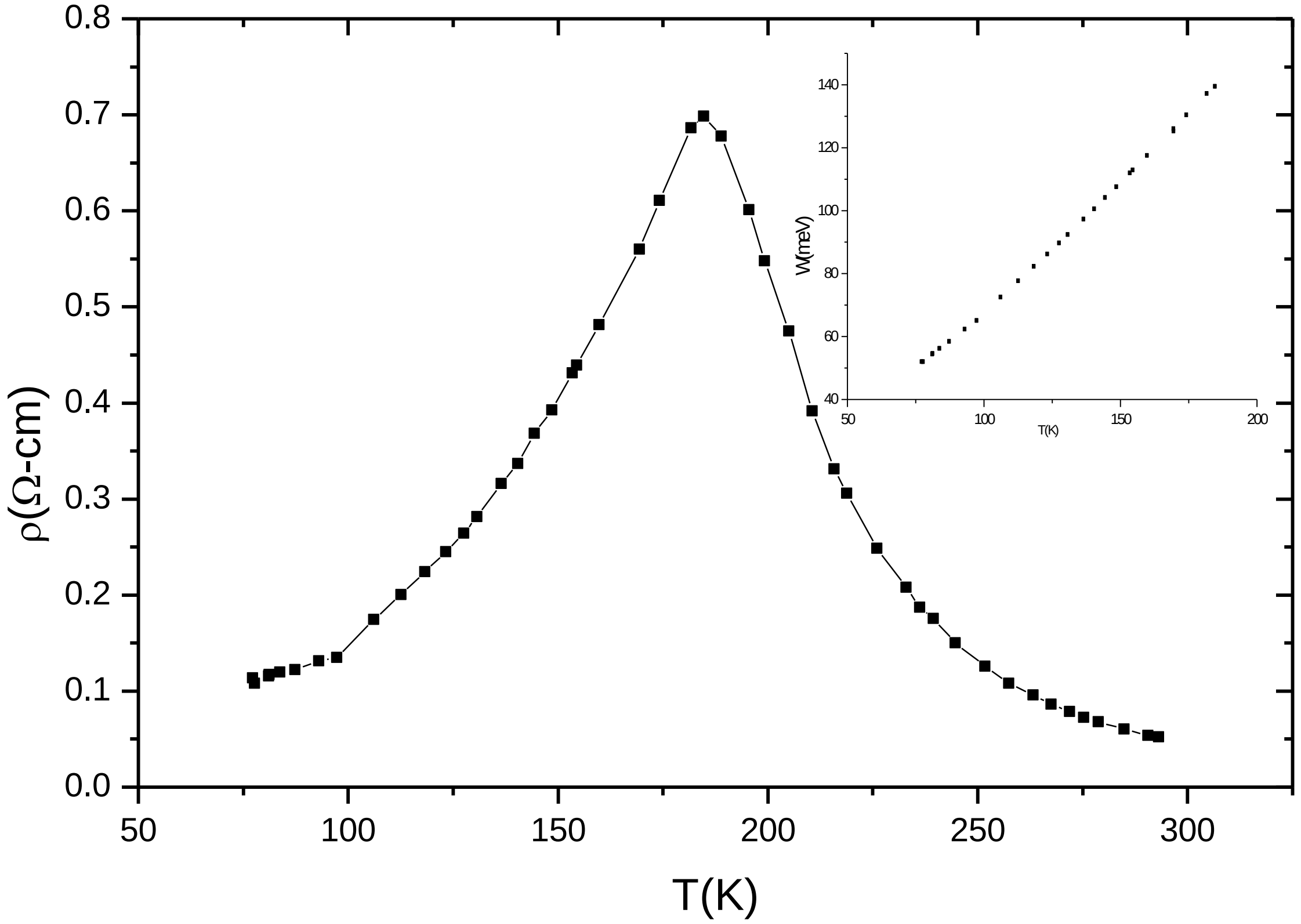

**Fig. 2**. The fig shows the variation of resistivity with temperature for $La_{0.65}Ca_{0.35}Mn_{0.95}Fe_{0.05}O_3$. The inset shows the variation of activation energy W (meV) with temperature for the same composition below $T_p$.

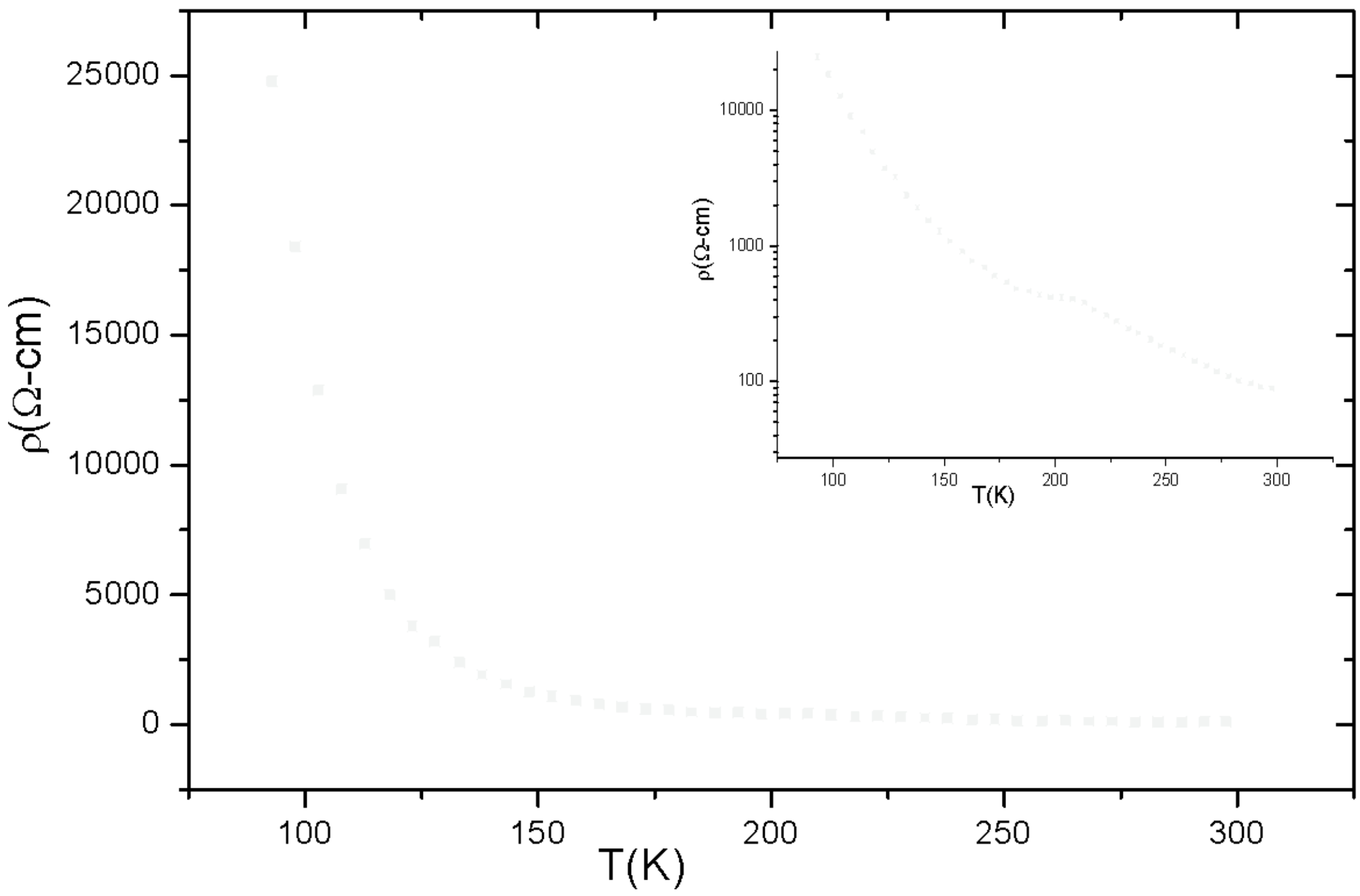

**Fig. 3**. The data shows the variation of resistivity with temperature for $La_{0.65}Pb_{0.35}Mn_{0.95}Fe_{0.05}O_3$ compound, large increase in resistivity below 120 K is observed. There is no transition from metal to insulator and remains insulating in the entire temperature region. The inset shows the behavior of resistivity for the same composition on log scale the flat region 210 K to 190 K is observed.

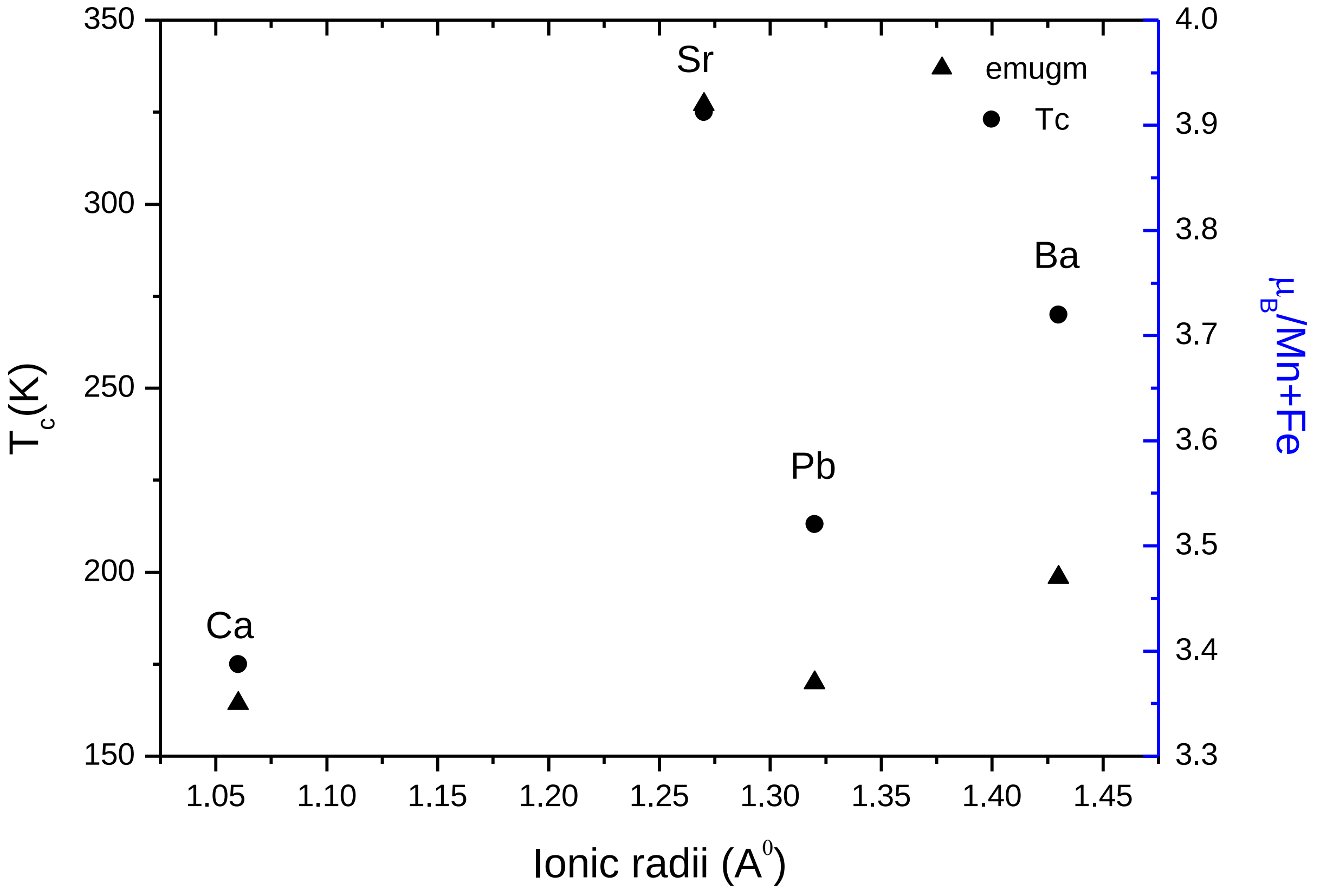

**Fig. 4**. Variation of magnetic moment ($\mu_B$/Mn+Fe) at 77 K and transition temperature $T_c$ as a function of ionic radii of divalent cations substituted on La site. The symbol ● shows the variation of transition temperature $T_c$ whereas the symbol ▲ shows the variation of magnetic moment at 77 K.

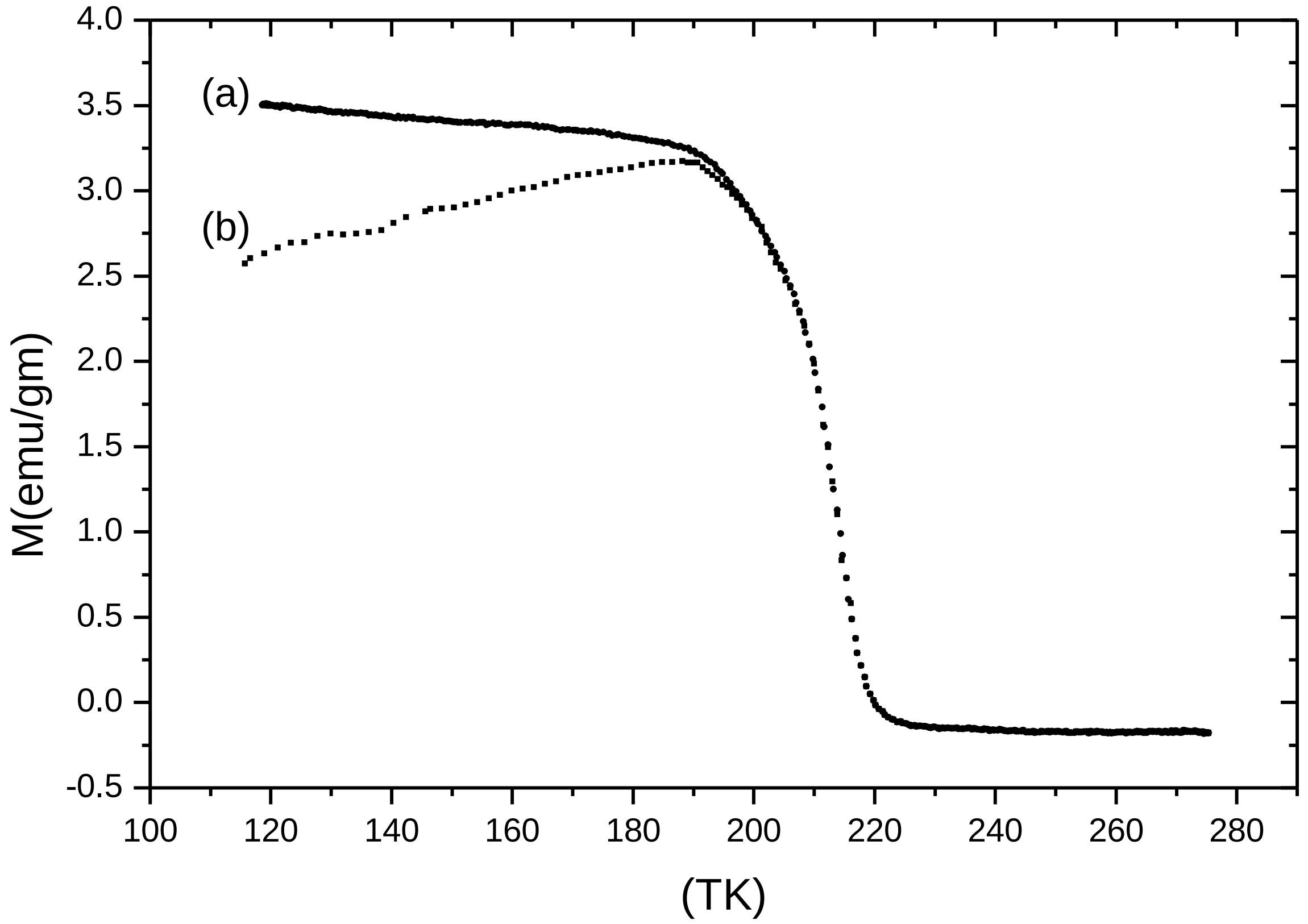

**Fig. 5**. The figure shows Magnetization (emu/gm) as a function of temperature for $La_{0.65}Pb_{0.35}Mn_{0.95}Fe_{0.05}O_3$. The cure (a) shows the variation of magnetization in ZFC mode and curve (b) shows the behavior of the same sample in FC mode. The decrease of magnetic moment below 110 K is observable.

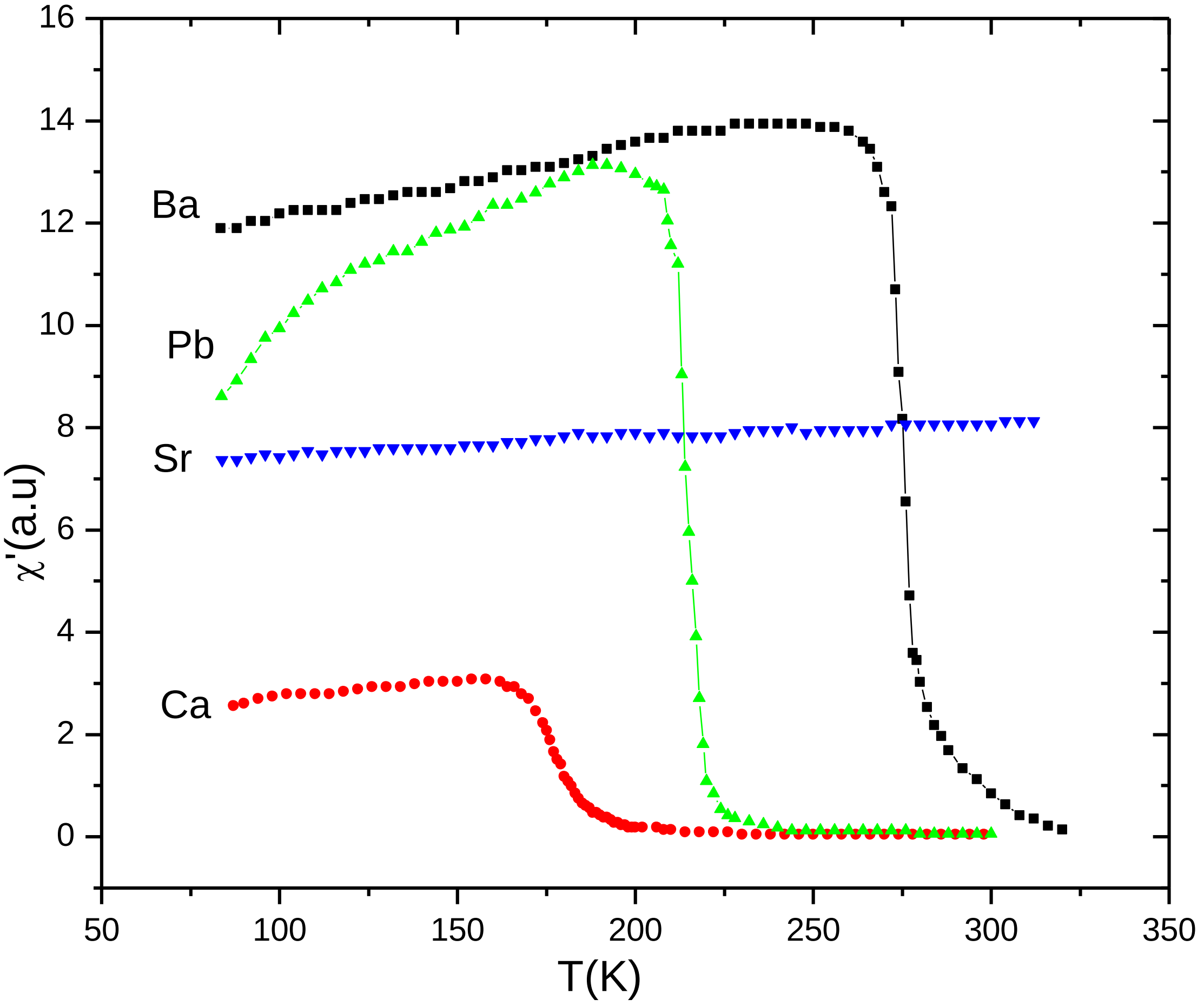

**Fig. 6:** Variation of in-phase part of the AC susceptibility with temperature for $La_{0.65}A_{0.35}Mn_{0.95}Fe_{0.05}O_3$ at a frequency of 131 Hz and AC amplitude of 5 Oe. The curves from top to bottom shows the data for Ba, Pb, Sr, and Ca doped composition. The Sr based sample has ferromagnetic throughout the temperature range studied here. For the other dopants one can easily find out $T_c$ from the curves of the susceptibility.

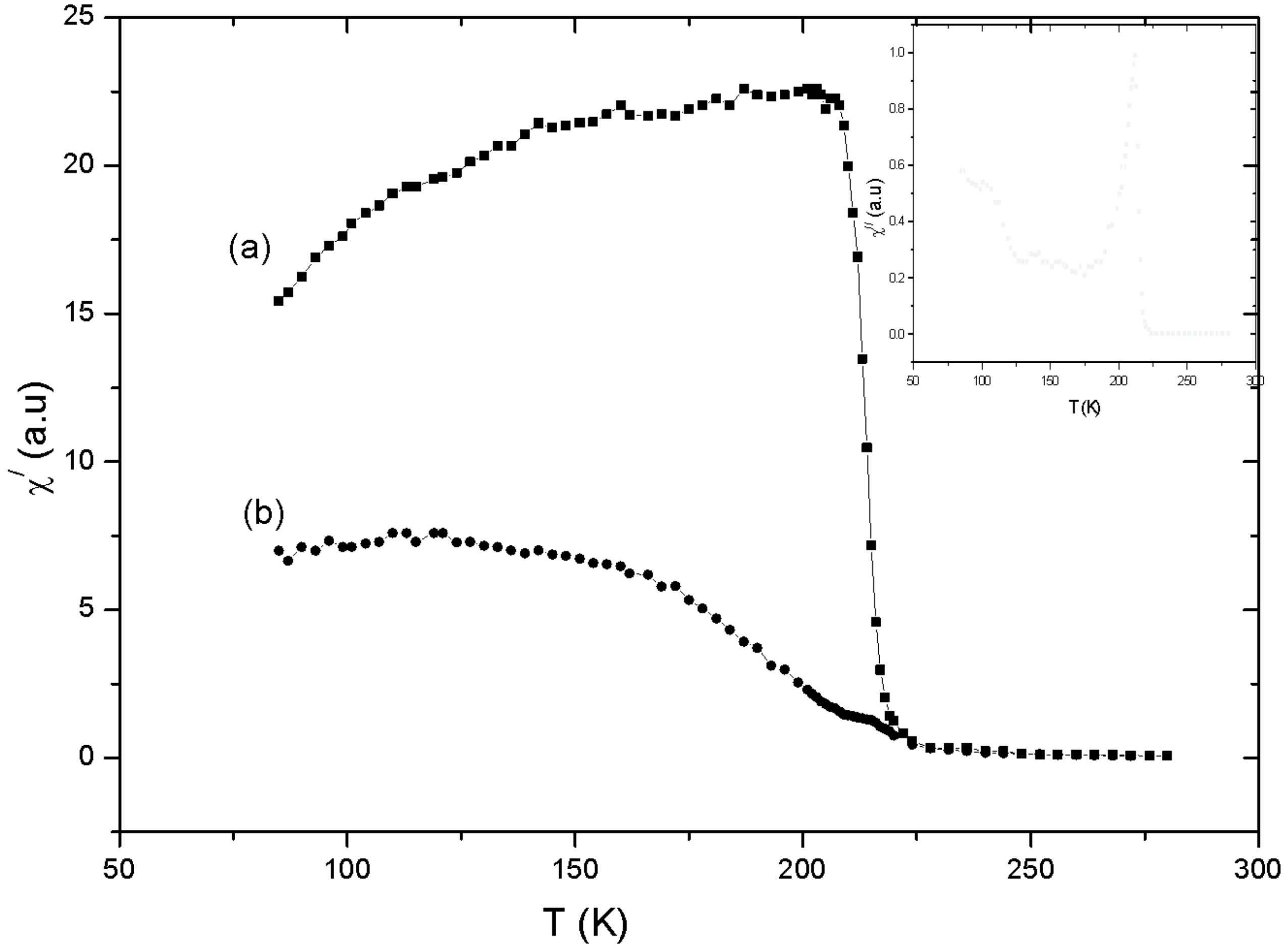

**Fig. 7:** Variation of AC susceptibility with temperature for $La_{0.65}Pb_{0.35}Mn_{0.95}Fe_{0.05}O_3$. The curve (a) shows the in-phase part of AC susceptibility in a frequency of 131 Hz and AC amplitude of 5 Oe. The curve (b) shows the in-phase part of the AC susceptibility in the superimpose DC magnetic field of 550 Oe. The inset shows the out of phase part ($\chi^{//}$) of AC susceptibility. The spin freezing behavior can be easily seen below 120 K.